\documentclass[a4paper,fleqn,final]{cas-dc}

\usepackage[T1]{fontenc}
\usepackage[utf8]{inputenc}
\usepackage[main=english,romanian]{babel}
\usepackage{textcomp}
\usepackage{hyperref}
\usepackage{amsmath}
\usepackage{amssymb}
\usepackage{graphicx}
\usepackage{booktabs}
\usepackage{tabularx}
\usepackage{threeparttable}
\usepackage[authoryear]{natbib}
\usepackage{longtable}
\usepackage{adjustbox}
\usepackage{caption}
\usepackage{etoolbox}
\usepackage{placeins}

\AtBeginEnvironment{table}{\let\sffamily\rmfamily}
\AtBeginEnvironment{figure}{\let\sffamily\rmfamily}
\AtBeginEnvironment{table*}{\let\sffamily\rmfamily}
\AtBeginEnvironment{figure*}{\let\sffamily\rmfamily}

\ExplSyntaxOn
\RenewDocumentCommand \firstname {}
	{ \textcolor{black}{\seq_use:Nn \l_stm_au_seq { ~ }} }

\cs_set:Npn \__first_footerline:
{
	\group_begin:
	\small
	\normalfont
	\ifnum\theblind>0\relax
	\else
	\__short_authors: :~
	\fi
	\itshape Preprint~submitted~to~ASABE~2026~AIM
	\group_end:
}

\RenewDocumentCommand \printorcid { } { }

\cs_set:Npn \__first_head:
{
	\parbox[t]{\textwidth}
	{
		\rule{\textwidth}{0pt}
	}
}

\cs_set:Npn \__cas_head:
{
	\parbox{\textwidth}
	{
		\rule{\textwidth}{0pt}
	}
}

\cs_set:Npn \__cas_foot:
{
	\parbox[t]{\textwidth}
	{
	 \rule{\textwidth}{.2pt}\\
	 \small
	 \normalfont
	 \__first_footerline:
	 \hfill Page~\thepage {}~of~ \lastpage
	}
}
\ExplSyntaxOff

\makeatletter
\ps@cas
\makeatother

\begin{document}

\shortauthors{Narimani et al.}
\shorttitle{AlphaEarth for Tomato Mapping}

\title[mode=title]{Mapping Tomato Cropping Systems in California Using AlphaEarth Geospatial Embeddings and Deep Learning Analysis}

\author[1]{Mohammadreza Narimani}
\ead{mnarimani@ucdavis.edu}

\author[1]{Alireza Pourreza}
\cormark[1]
\ead{apourreza@ucdavis.edu}

\author[1]{Parastoo Farajpoor}
\ead{pfarajpoor@ucdavis.edu}

\affiliation[1]{organization={Department of Biological and Agricultural Engineering, University of California, Davis},city={Davis},state={CA},postcode={95616},country={USA}}

\cortext[1]{Corresponding author}

\begin{abstract}
Field-scale crop maps support supply-chain forecasting and policy, yet statewide crop identification still often depends on retrospective surveys or remote-sensing workflows built around hand-engineered spectral features. Those pipelines can be accurate, but they require repeated preprocessing and often lose robustness across years. This study evaluated whether Google DeepMind's AlphaEarth geospatial embeddings can serve as an analysis-ready alternative for mapping processing tomato systems in California. LandIQ 2018 crop polygons were used to assemble a balanced reference dataset of 4,742 tomato and 4,742 non-tomato fields. For each polygon, 64-band AlphaEarth embedding chips were extracted and aligned with binary masks, then divided into spatially independent training (n\,=\,6,638), validation (n\,=\,1,422), and test (n\,=\,1,424) sets. A U-Net segmentation model was trained on AWS SageMaker using a composite masked binary cross-entropy and soft Dice loss. To complement hard predictions, Monte Carlo dropout was retained at inference and repeated 100 times per chip to estimate predictive mean and variance. On the independent test set, the model achieved 99.19\% pixel accuracy, 98.69\% precision, 99.40\% recall, 99.04\% F1 score, 98.11\% intersection over union, and 99.02\% chip accuracy. Uncertainty maps were consistently highest near field edges and low within field interiors. The results show that AlphaEarth embeddings retain crop-relevant spatial and temporal structure and can support accurate, field-scale tomato mapping without manual feature engineering.
\end{abstract}

\begin{keywords}
AlphaEarth \sep crop mapping \sep deep learning \sep processing tomatoes \sep segmentation \sep satellite embedding \sep uncertainty quantification
\end{keywords}

\maketitle


\section{Introduction}\label{sec:introduction}

Accurate crop mapping underpins a wide range of decisions in modern agriculture, from acreage estimation and irrigation planning to market forecasting and policy design \citep{Badshah2024,Hudait2022}. Research in this area spans multiple observation scales. At the plant scale, leaf-level spectroscopy and imaging are used to characterize physiology, disease symptoms, and varietal traits \citep{Cozzolino2016,Farajpoor2025b}. At the field scale, unmanned aerial systems resolve canopy patterns and within-field heterogeneity at high spatial detail \citep{Gano2024,Narimani2024}. Recent trait-to-spectra modeling work also shows how data-driven neural networks can link measured leaf traits with detailed reflectance patterns, reinforcing the importance of plant-level spectral information for remote sensing applications \citep{Farajpoor2025a}. At the regional scale, satellite remote sensing remains the most practical option for repeated crop inventories across large production areas, although the gain in coverage is usually accompanied by mixed pixels, cloud contamination, sensor differences, and a heavier preprocessing burden \citep{Narimani2025a,Zhang2020}.

These challenges are especially important for processing tomatoes in California, where field-level maps have value beyond simple acreage reporting. Reliable tomato maps can support water accounting, yield forecasting, harvest logistics, contract planning, and regional policy evaluation \citep{Brodt2013}. California is the dominant production region for processing tomatoes in the United States, so classification errors at statewide scale can propagate into both economic and resource-management decisions \citep{Espinoza2023,Narimani2025b}.

Operational crop mapping products such as LandIQ and USDA layers are highly valuable because they provide field boundaries and class labels at a large scale \citep{Espinoza2023,Lark2021}. However, those products are retrospective by design and depend on repeated cycles of imagery interpretation, survey information, and model updating. In parallel, many research pipelines still rely on handcrafted spectral indices or manually designed temporal features derived from Landsat, Sentinel-2, or related sensors. Recent reviews of Sentinel-2-based yield estimation highlight similar trends, showing that machine learning, deep learning, crop-model assimilation, and optical--SAR data fusion are increasingly used for field-level agricultural monitoring, while cloud gaps, limited ground truth, and transferability across years and regions remain persistent challenges \citep{Narimani2026}. Those approaches can perform well, but they often require cloud masking, atmospheric harmonization, gap filling, and case-specific tuning, and their transfer across years is not always reliable when crop rotation, weather, or management practices change \citep{Hu2019,Vidican2023}.

Recent geospatial foundation models offer a different starting point. Google DeepMind's AlphaEarth Foundations produces analysis-ready annual embeddings in which each 10\,m pixel is represented by a 64-dimensional vector that summarizes information from multi-source Earth observation data. This representation can be viewed as a data-driven form of dimensionality reduction: like principal component analysis, it compresses a high-dimensional signal into a smaller feature space, but it does so with a nonlinear, multimodal model trained on large spatiotemporal datasets rather than on a linear projection. The individual embedding axes are not intended to be interpreted on their own; instead, the 64-dimensional pattern is treated as a joint descriptor of land-surface behavior \citep{Brown2025,Ma2026}.

This study investigated whether AlphaEarth embeddings can separate processing tomato fields from a balanced and diverse non-tomato class across California. Using LandIQ 2018 polygons as reference data, we built a field-scale segmentation dataset, trained a 64-channel U-Net model, and evaluated both predictive performance and spatial uncertainty on geographically independent test fields. Although the reported supervision is anchored to 2018 labels, the data pipeline was designed around the annual AlphaEarth stack, making the framework suitable for multi-year expansion in future work.

\begin{figure*}[t]
\centering
\includegraphics[width=0.85\textwidth]{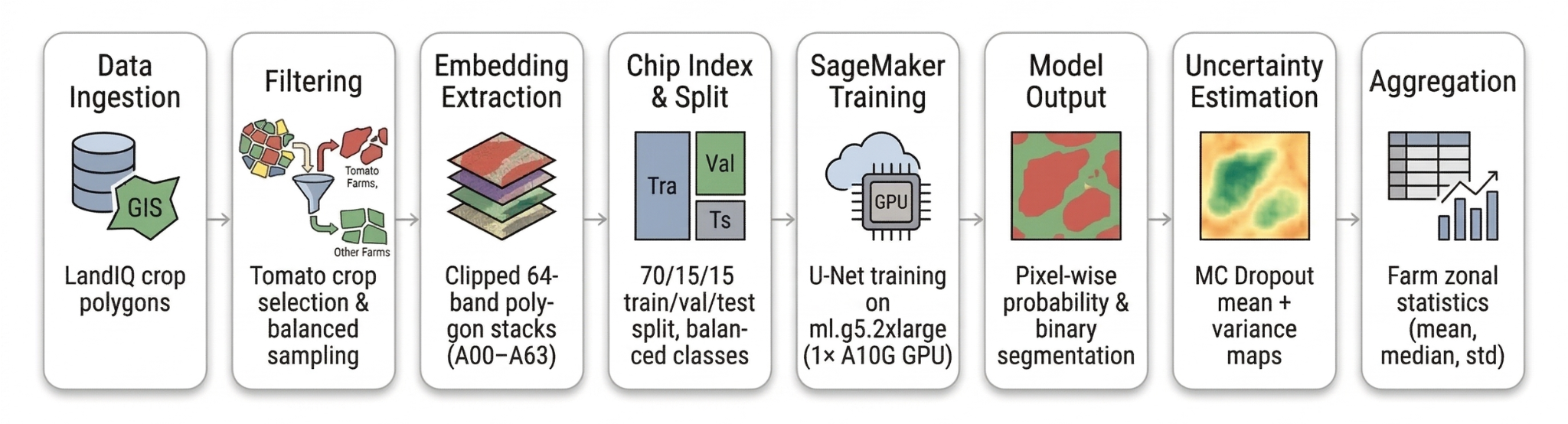}
\caption{End-to-end workflow used to build the dataset, train the segmentation model, generate pixel-wise probability maps, estimate uncertainty, and summarize field-level outputs.}
\label{fig:workflow}
\end{figure*}

\section{Methodology}\label{sec:methodology}

\subsection{Study Area and Data Generation}

The study focused on California agricultural landscapes represented in the LandIQ 2018 statewide crop mapping layer. LandIQ provides field-scale polygons for crop and land-use classes across the state and reports high statewide mapping accuracy. From this layer, 4,742 polygons labeled as tomato were extracted to define the positive class. A balanced negative class of 4,742 non-tomato polygons was then assembled from other agricultural categories, including alfalfa, wheat, corn, beans, orchards, and additional field crops. Rather than defining the negative class as a single alternative crop, we constructed it from a balanced mix of non-tomato crop groups selected to represent a broad cross-section of California agriculture. Although this class does not capture every crop grown in the state, it was designed to include diverse major crop types and therefore required the model to learn tomato-specific patterns rather than a simple contrast between tomatoes and one competing crop.

The resulting dataset contained 9,484 field polygons. For each polygon, a binary mask was generated so that pixels inside tomato polygons were labeled as tomato and pixels inside non-tomato polygons were labeled as non-tomato. AlphaEarth embedding layers were clipped to the polygon footprints, and aligned raster chips were created locally before being stored in Amazon S3 for model training and inference. Chips affected by NaN or NoData margins were masked so that only valid agricultural pixels contributed to loss calculation and metric reporting.

To ensure an independent evaluation, the polygons were partitioned spatially before model development so that nearby fields did not appear across the training, validation, and test sets. The final split contained 6,638 training polygons, 1,422 validation polygons, and 1,424 test polygons. This design ensured that the reported test metrics reflected performance on unseen locations rather than memorization of nearby fields with similar geometry or management.

\subsection{AlphaEarth Geospatial Embeddings}

Input features were taken from the public AlphaEarth Satellite Embedding dataset, which provides annual 64-band images (A00--A63) at 10\,m spatial resolution. Each image summarizes the multi-sensor surface trajectory of a single calendar year rather than a seasonal or near-real-time interval. The public Earth Engine release is organized as one annual layer per year, with coverage currently available from 2017 through 2024; accordingly, AlphaEarth should be treated as a year-level analysis product that becomes available only after annual observations have been assembled and processed. Each band is dimensionless and bounded between $-1$ and $1$, but the bands should be interpreted collectively as coordinates in a learned embedding space rather than as standalone physical measurements \citep{Brown2025}.

For this study, each field chip was represented as a tensor $\mathbf{X} \in \mathbb{R}^{64 \times H \times W}$, where $H$ and $W$ denote the rasterized chip dimensions. Because the embeddings are learned from multi-sensor observations and temporal context, they implicitly encode patterns that would otherwise require extensive feature engineering from raw optical or radar imagery. Pseudo-RGB visualizations were generated only for figure display by assigning three embedding axes to red, green, and blue channels; the model itself used all 64 channels simultaneously.

\subsection{Model Architecture and Training}

\begin{figure*}[t]
\centering
\includegraphics[width=0.85\textwidth]{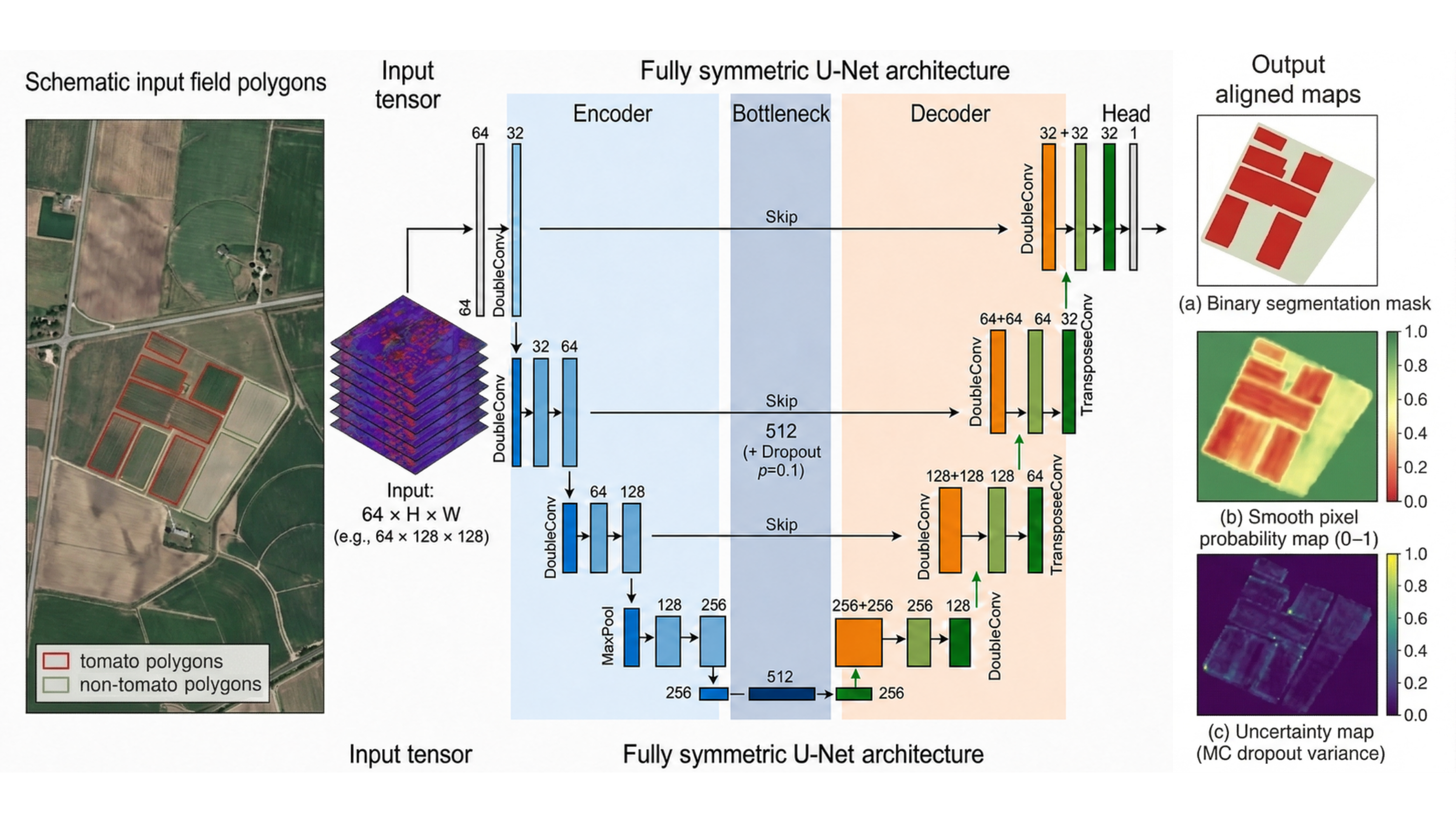}
\caption{U-Net architecture used to map tomato versus non-tomato pixels from 64-band AlphaEarth embedding chips and to estimate predictive uncertainty through Monte Carlo dropout.}
\label{fig:unet}
\end{figure*}

A U-Net segmentation model was adopted because it couples local boundary detection with broader spatial context through encoder-decoder skip connections \citep{Ronneberger2015}. The network accepted 64 input channels directly from the AlphaEarth tensor stack and used a base width of 32 feature maps. The model output a dense probability map aligned to the input chip, allowing tomato detection to be treated as a pixel-wise binary segmentation task rather than as a single field label. Figure~\ref{fig:unet} summarizes the architecture and the relationship between the embedding input, segmentation output, and uncertainty output.

For each valid pixel $i$ with ground-truth label $y_{i} \in \{0, 1\}$, raw model output $z_{i}$ was converted to probability $p_{i}$ using a sigmoid function \citep{Kyurkchiev2015}:
\begin{equation}
p_{i} = \sigma(z_{i}) = \frac{1}{1 + e^{-z_{i}}}
\end{equation}
The loss function combined masked binary cross-entropy and soft Dice loss to balance pixel-wise discrimination with boundary overlap. Let $m_{i}$ denote the valid-pixel mask and $\varepsilon$ a small constant for numerical stability. The total objective was defined as the sum of the two losses \citep{Azad2023}:
\begin{equation}
\mathcal{L}_{BCE} = -\frac{1}{\sum_{i}^{} m_{i}} \sum_{i}^{} m_{i} \left[ y_{i} \log(p_{i}) + (1 - y_{i}) \log(1 - p_{i}) \right]
\end{equation}
\begin{equation}
\mathcal{L}_{Dice} = 1 - \frac{2 \sum_{i}^{} m_{i} p_{i} y_{i} + \varepsilon}{\sum_{i}^{} m_{i} p_{i} + \sum_{i}^{} m_{i} y_{i} + \varepsilon}
\end{equation}
\begin{equation}
\mathcal{L} = \mathcal{L}_{BCE} + \mathcal{L}_{Dice}
\end{equation}
Training was performed on AWS SageMaker using an ml.g5.2xlarge instance equipped with one NVIDIA A10G GPU. Automatic mixed precision was enabled to improve memory efficiency and training throughput. The optimizer used a learning rate of 0.001, weight decay of 0.0001, batch size of 24, and a 30-epoch training schedule. Validation loss, pixel accuracy, precision, recall, F1 score, and intersection over union were tracked at each epoch on valid pixels only.

\subsection{Uncertainty Quantification}

A single deterministic prediction is useful for mapping, but it does not reveal where the model is uncertain. To estimate predictive confidence, dropout layers were kept active during inference following the Monte Carlo dropout framework \citep{Gal2015}. Each chip was passed through the network 100 times, producing a set of stochastic probability maps $\{p^{(t)}\}_{t=1}^{T}$ with $T = 100$. The predictive mean served as the final soft classification map, while the pixel-wise predictive variance was interpreted as spatial uncertainty. In practice, high variance was expected near field borders, mixed pixels, and geometrically ambiguous edges, whereas low variance indicated stable agreement across repeated forward passes.

\begin{equation}
\widehat{\mu} = \frac{1}{T} \sum_{t=1}^{T} p^{(t)}
\end{equation}
\begin{equation}
\widehat{\sigma}^{2} = \frac{1}{T} \sum_{t=1}^{T} \left( p^{(t)} - \widehat{\mu} \right)^{2}
\end{equation}

\section{Results}\label{sec:results}

The model produced highly consistent results on the spatially independent test set of 1,424 fields. Table~\ref{tab:performance} summarizes the quantitative evaluation. Pixel accuracy reached 99.19\%, with precision of 98.69\% and recall of 99.40\%, yielding an F1 score of 99.04\%. Intersection over union was 98.11\%, and chip accuracy was 99.02\%. Taken together, these values indicate that the network was not simply identifying large patches of cultivated land; it was also tracing tomato field boundaries with very limited confusion against the negative class.

\begin{table}[ht]
\centering
\caption{Test-set performance of the tomato segmentation model on 1,424 spatially independent fields.}
\label{tab:performance}
\begin{tabular*}{\columnwidth}{@{\extracolsep{\fill}}lr@{}}
\toprule
\textbf{Metric} & \textbf{Score (\%)} \\
\midrule
Pixel Accuracy & 99.19 \\
Precision & 98.69 \\
Recall & 99.40 \\
F1 Score & 99.04 \\
Intersection over Union (IoU) & 98.11 \\
Chip Accuracy & 99.02 \\
\bottomrule
\end{tabular*}

\vspace{1ex}
\raggedright\footnotesize
\textit{Note:} Metrics were computed on the spatially independent test set using valid pixels only.
\end{table}

\begin{figure*}[t]
\centering
\includegraphics[width=0.85\textwidth]{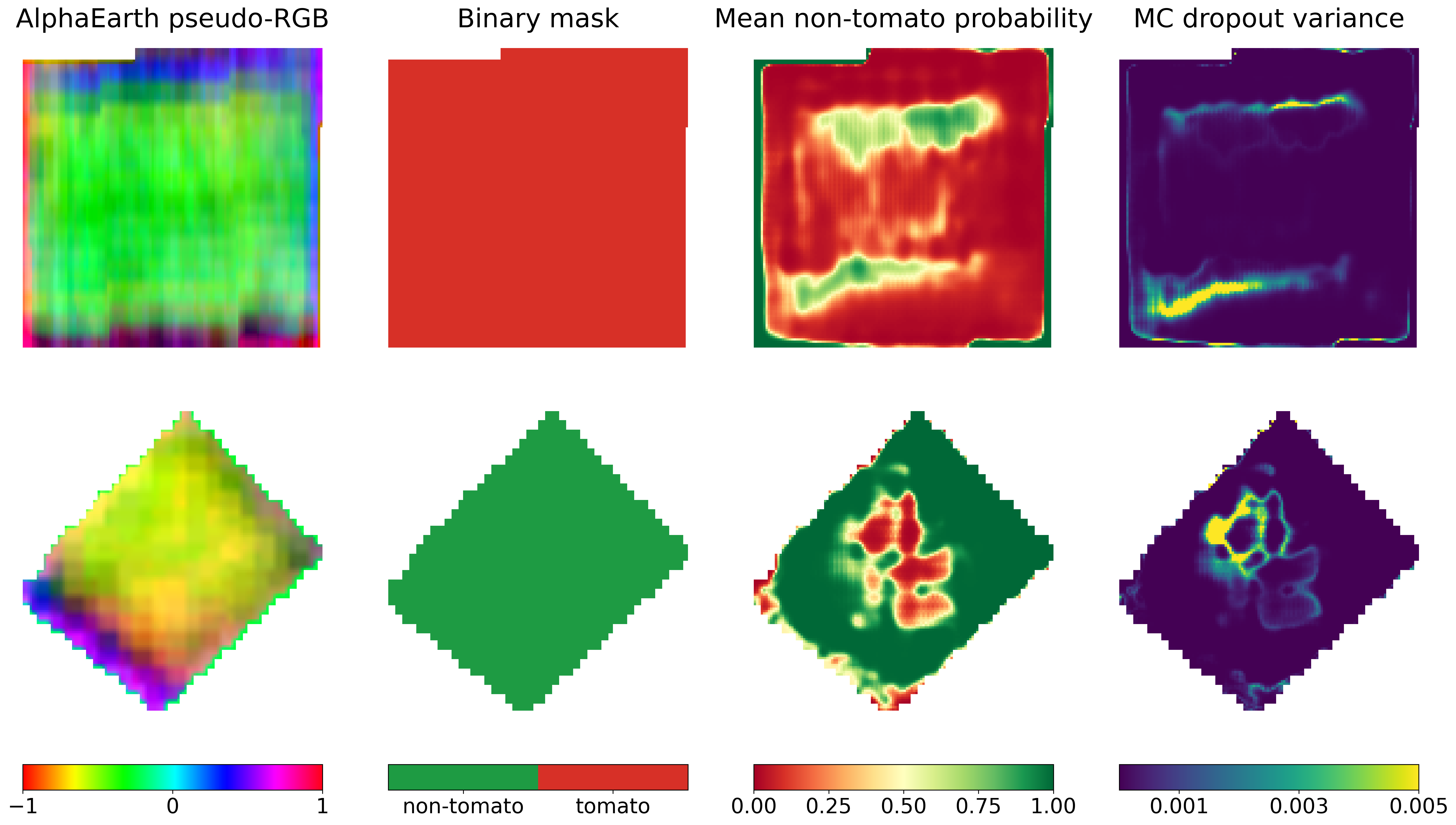}
\caption{Representative prediction results for one tomato field (top row) and one non-tomato field (bottom row). Columns show pseudo-RGB embedding visualization, reference mask, mean predicted non-tomato probability, and Monte Carlo dropout variance.}
\label{fig:results}
\end{figure*}

Qualitative examples in Figure~\ref{fig:results} support the numerical results. The top row shows a representative tomato field, with low predicted non-tomato probability inside the reference polygon, whereas the bottom row shows a representative non-tomato field with the opposite response. The probability maps are spatially smooth within field interiors and change most strongly near crop edges, roads, or partially mixed margins. The uncertainty maps reveal a consistent spatial structure: variance is generally low across the interior of well-defined fields and increases around borders, narrow protrusions, and irregular boundaries. This behavior is desirable because those locations are most sensitive to mixed pixels, rasterization effects, and subtle class transitions. Importantly, the model remained stable even when the negative class included visually similar annual crops such as wheat, alfalfa, or other intensively managed fields. That finding suggests the embedding space retains temporal and structural cues that extend beyond simple color or texture differences in a single image.

\section{Discussion and Conclusion}\label{sec:discussion}

The strong performance can be understood in light of what AlphaEarth represents. Rather than starting from a single-date optical scene, the model ingests a compact summary of multi-sensor and temporal information. Processing tomato fields have distinctive growth cycles, canopy development, harvest timing, field geometry, and surrounding management context. Those signals are difficult to capture with one or two hand-engineered indices, but they can be preserved in a learned embedding that has already integrated information across sensors and time.

That feature space changes the role of the downstream model. Instead of learning directly from raw reflectance values that still contain clouds, seasonal gaps, or cross-sensor inconsistencies, the U-Net operates on analysis-ready descriptors. The segmentation network can therefore focus on separating tomato from non-tomato patterns and refining boundaries. The uncertainty maps add a second practical benefit: they indicate where a map is likely reliable and where human review or auxiliary data may still be useful.

Several limitations should be noted. First, the reference labels were drawn from LandIQ 2018, so the reported supervised results reflect one labeled year even though the data pipeline was built around annual AlphaEarth layers. A stronger assessment of temporal robustness will require matched labels across multiple years. Second, the negative class was intentionally balanced, which is useful for controlled training but does not fully reproduce the class imbalance present in statewide operational mapping. Third, Monte Carlo dropout improves interpretability but increases inference cost because each chip must be evaluated repeatedly. This trade-off may matter at statewide or multi-country scale unless the workflow is further parallelized.

Even with those limitations, the results show that AlphaEarth embeddings can support highly accurate field-scale tomato mapping in California when coupled with a relatively compact segmentation model. The approach offers a practical alternative to pipelines built around manually engineered spectral features. Future work should extend the label base across multiple years, broaden the non-tomato inventory to additional crop systems, test transferability beyond California, and examine whether the same embedding space can support more detailed agricultural tasks such as stress detection, disease monitoring, and management-zone analysis. In that sense, this study serves not only as a tomato classification experiment, but also as an early demonstration of how geospatial foundation-model embeddings can be turned into operational agricultural layers.

\bibliographystyle{cas-model2-names}
\bibliography{references}

\end{document}